\documentstyle[12pt]{article}
\begin{document}
\def\Ket#1{|#1 \rangle}
\def\PP{\vskip.1in}
\def\mod{{\rm mod \,} }
\centerline{{\bf The Prime Factorization Property of Entangled Quantum States}}
\vskip.15in
\centerline{Daniel I. Fivel}
\vskip.15in
\centerline{ Department of Physics }
\centerline{ University of Maryland, College Park, MD 20742-4111 }
\vskip.2in
\centerline{{\bf Abstract}}
\PP
Completely entangled quantum states are shown to factorize into tensor products
of entangled states whose dimensions are powers of prime numbers. The entangled
states
of each prime-power dimension transform among themselves under a finite
Heisenberg group. We
 consider processes in which factors are exchanged between entangled
states and study canonical ensembles in which these processes occur. It is
shown that the Riemann zeta function is the appropriate partition function
and that the Riemann hypothesis makes a prediction about the high temperature
contribution of modes of large dimension.    \PP
\PP
\PP
Completely entangled quantum states are of considerable interest not only in
the foundations of
quantum mechanics\cite{FIVa,FIVb}, but also because of the possibility of their
use in secure
communication schemes\cite{BEN,WOOT}. It has been shown\cite{FIVc} that any
completely entangled
two-particle state  is expressible in the following form: Let $\Ket{n,\nu},\;
n=1,2,\cdots, N$ be any basis of a Hilbert space of dimension $N$. Here $\nu =
1,2$ is usually taken
to be a particle label, but more generally it may label any pair of Hilbert
 spaces of
the same dimension. Let ${\cal U}$ indicate an arbitrary anti-unitary
 transformation on the
N-dimensional Hilbert space and define
$$\Ket{n^{\cal U},\nu} = {\cal U}\Ket{n,\nu}.\eqno(1)$$
Then the state
$$\Ket{{\cal U}} \equiv N^{-1/2}\sum_{n=1}^{N}{\Ket{n,1}\otimes\Ket{n^{{\cal
U}},2}}.
 \eqno(2) $$
is a completely entangled state. In this state  there is equal likelihood of
finding
particle-1 in any state $\Ket{x,1}$, but it is certain that it will be found in
this state if its
partner is found in the state $\Ket{x^{{\cal U}},2}$.  Note that our use of the
same letter ${\cal
U}$ to label both the entangled state and the operator that appears on the
right is justified by the
fact that
 the right side is independent
of the choice of basis. This fact (which makes essential use of the
anti-unitarity) is
what produces the perfect correlation for any choice of $\Ket{x,1}$.\PP

There is no way to write an entangled state $\Ket{{\cal U}}$ in the form of a
monomial tensor product of a state in the $\nu = 1$ space with a state in the
$\nu = 2$ space.
However, let us take note of a structure that can be illustrated by the
following example:
 Suppose
$N=6$ and the the six states $n=1,\cdots,6$ correspond to the values of the
pair $n=(n',n'')$ where
$n' = 1,2$ indexes a two dimensional Hilbert space of ``spin", and $n''=1,2,3$
a three dimensional
space of ``color". Then each state $\Ket{n,\nu}$ can be written in the form:
$$
\Ket{n,\nu} = \Ket{n',\nu}\otimes \Ket{n'',\nu}.
\eqno(3)
$$
If ${\cal U}$ is expressible in the form:
$$
{\cal U} = {\cal S} \otimes {\cal C}
\eqno(4)
$$
where the operators on the right are anti-unitary transformations in the spin
and color spaces
respectively we have:
$$
\Ket{{\cal U}}  = \Ket{{\cal S}}\otimes\Ket{{\cal C}},
\eqno(5)
$$
with
$$\Ket{{\cal S}} = 2^{-1/2}\sum_{n'=1}^{2}{\Ket{n',1}\otimes\Ket{n'^{{\cal
S}},2}},$$

 $$\Ket{{\cal C}} = 3^{-1/2}\sum_{n''=1}^{3}{\Ket{n'',1}\otimes\Ket{n''^{{\cal
C}},2}}.
 \eqno(6) $$

Thus while the state $\Ket{{\cal U}}$ does not factorize with respect to the
particle label $\nu$, it
{\it does} factorize with respect to the two {\it properties}. Moreover {\it
the factorization is
into a tensor product of two states that are also completely entangled states.}
Examples of the same
sort  can obviously be produced for arbitrary {\it non-prime} $N$ and any
number of factors.\PP

If $N$ is a non-prime the question then arises as to when the Hilbert space
decomposes into
a tensor product.
Such a factorization implies that all linear (or anti-linear) transformations
are
expressible as tensor products of operators belonging to the factors, i.e.\ we
are only allowed to
consider operations in which the factors transform among themselves with no
interference
between distinct factors. Conversely, if all physically implementable
transformations in the system
of interest can be constructed from tensor products then the space will be
representable as a tensor
product.\PP

Thus to permit factorization we must show that there is a set of operators that
is large enough
to characterize all interesting physical processes, but small enough to be
expressible as tensor
products of operators associated with the factor dimensions. Since we confine
ourselves to
entangled states, we need only consider transformations that take entangled
states into one
another. In particular if we restrict to a complete set of entangled states of
any dimension $N$, we
need only consider operators that transform the set among one another and these
form a {\it finite}
group. The structure of such transformations has been previously analyzed by
the author
\cite{FIVc}, and is as follows:
\PP
The Heisenberg group ${\cal G}_N$ is defined abstractly as the group generated
by
$\sigma,\tau,I$ where $I$ is the identity and
$$
\sigma \tau = \omega\tau\sigma,\;\; \omega = e^{2\pi i/N}.
\eqno(7)
$$
It can be represented in a Hilbert space ${\cal H}$ of dimension $N$ with basis
$\Ket{j}, \; j= 0,1,\cdots {\mod N}$ by:
$$
\sigma\Ket{j} = \omega^j\Ket{j},\; \tau\Ket{j} = \Ket{j+1},\;\;
{\rm with}\; \tau\Ket{N-1} = \Ket{0}.
\eqno(8)
$$
The $N^2$ operators $\sigma^j\tau^k, j,k = 0,1,\cdots, N-1$ of ${\cal G}_N$
acting on either
particle cause the entangled state to hop from one point of an $N\times N$
lattice to another. This
lattice resembles a phase space in which $\sigma$ and $\tau$ act like a
coordinate and momentum
generator.
\PP
 In the case where $N$ is a prime $p$ it is  known \cite{FIVc} that ${\cal
G}_p$ is the only
group with $p^2$ elements that will cause a complete set of mutually orthogonal
entangled states to
transform  among themselves. More generally  finite groups that transform
sets of
$N^2$ linearly-independent entangled states of dimension $N$ among themselves
will be direct
products of Heisenberg groups ${\cal G}_{n_1}\times {\cal
G}_{n_2}\times\cdots\times {\cal G}_{n_j}$
where $n_1n_2\cdots n_j = N$ and each factor is a {\it prime power}. To
understand why it is
possible that a Heisenberg group of prime-power order may not decompose further
one need only
compare the groups
 ${\cal G}_{p^2}$ and ${\cal G}_{p}\times {\cal G}_{p}$ both of which have the
same number of
elements. In the construction of ${\cal G}_N$ one needs the $N$'th roots of
unity (see (7) above).
If $N = pq$ with distinct primes $p,q$ then every $N$'th root of unity is a
product of some
$p$'th and some $q$'th root of unity, a property that disappears if $p = q$.
\PP

We see then that the Heisenberg group
 supplies us with a set of operations sufficiently large to transform a
complete system of
entangled states into one another, and, moreover, is represented in the tensor
product Hilbert
space. Thus we are able to describe the physics of a complete system of
entangled states of
non-prime dimension via the tensor product of entangled states whose dimensions
are prime power
factors of $N$.\PP

The factorization property naturally suggests that we consider ensembles of
entangled states
which interact by exchanging factors, e.g.\ an $N_1 = 6$ state
interacting with an $N_2 = 4$ state to form an $N_3=12$ and an $N_4=2$ state.
In such interactions
the sum of the logarithms of the dimensions is the {\it additive} conserved
quantity. Hence the
partition function of a canonical ensemble will be:
 $$
{\cal Z} = \sum_{N=1}^{\infty}{e^{-\beta \log N}} =
\sum_{N=1}^{\infty}{N^{-\beta}} = \zeta(\beta)
\eqno(9)
$$
which is the Riemann zeta function. Using the Euler product formula
$$
\zeta(\beta) = \prod_{primes}{(1 - p^{-\beta})^{-1}}
\eqno(10)
$$
one then computes the expectation value of $<\log N>$ in this ensemble to be
 $$
< \log N > = -(d/d\beta)\log\zeta(\beta) =
$$
$$
 \sum_{primes}{ \log p \cdot (p^{-\beta} + p^{-2\beta} + \cdots)} =
 \sum_{N=2}^{\infty}{ { {\Lambda (N)}\over{N^{\beta}}}}
 \eqno(11)
 $$
where
$$\Lambda(N)=\left\{ \matrix{\log p \;  \; \hbox{ if N is a {\rm a prime
power}}
\;\hfill\cr
  \hfill\cr
 0 \;\; {\rm otherwise}\hfill\cr} \right.
\eqno(12)
$$\PP
At low temperatures (large $\beta $) only the low values of $N$ contribute, and
in the extreme case
$T=0$ we have only $N = 1$ i.e.\ a completely factorized state. Thus
entanglement disappears at $T = 0$.\PP

The high temperature (low $\beta$) limit is more interesting. Here the large
values of $N$
contribute, and the partition function has a simple pole at $\beta =
1$ indicating a phase transition. A rigorous way to study the $\beta \to 0$
behavior is by analytic
continuation which involves one in a study of the zeros of the zeta function in
the
critical strip $0 < Re(\beta) < 1$. However, we can expose part of the physical
content of the high
temperature behavior in the following simpler manner: For any $x$ and $\beta$
consider the quantity:
$$
f_\beta(x) = x^{-1}\sum_{N=2}^{x}{\Lambda(N)/N^{\beta}},
\eqno(13)
$$
which measures the average contribution to the expectation value in (9) coming
from the modes
associated with $N \leq x$. For $\beta \to 1$ the series diverges only
logarithmically as $x\to
\infty$ so that the average contribution measured by $f$ still goes to zero
indicating that there is
not yet complete dominance of the high $N$ modes. This situation changes,
however, as $\beta \to 0$.
To see how this dominance develops we  examine the asymptotic behavior of:
 $$
\psi(x) \equiv \sum_{N=2}^{x}{\Lambda(N)}.
\eqno(14)
$$

Before doing so we remark on the significance of this function in number
theory: One
can show that the number $\pi(x)$ of primes  less than $x$ has asymptotic
behavior\cite{RAD} given by $\psi(x)/\log x$. In fact the usual proof of the
prime number theorem (which tells us the leading asymptotic behavor) is carried
out by showing that
$$
\psi(x) = x + r(x),\;  \; r(x) = o(x).
\eqno(15)
$$
Thus the corrections to the prime number theorem are obtained from a study of
the asymptotic
behavior of $r(x)$. The importance of the
Riemann hypothesis concerning the distribution of the zeros of the zeta
function derives from the
fact that if it is true one can derive the strongest possible asymptotic
estimate for $r(x)$ namely:
$$
r(x)/x = O(e^{-c\sqrt{\log x}}), \; \; c > 0.
\eqno(16)
$$
In the context of our analysis of the high temperature behavior, we see that
the Riemann hypothesis
implies that $f_o(x)$ defined by (13) has the asymptotic form:
$$
f_o(x) \approx 1 + O(e^{-c\sqrt{\log x}}),
\eqno(17)
$$
and thus makes a prediction about the asymptotic high temperature distribution
of
modes in ensembles of entangled states. \PP

In an earlier part of the discussion we took note of the fact that there is a
non-uniqueness in
the factorization of entangled states of dimension $N$ when any of the prime
factors occur to
powers higher than the first. It is thus of interest to consider ensembles in
which this does not
happen, i.e.\ in which only those integers are permitted whose prime factors
are at most of the
first power. Since the primes label the possible entanglement modes, one sees
that such a
restricted ensemble has a fermionic character, while with unrestricted $N$
it has a bosonic character. In the restricted case the partition function is
 expressed as the Euler product:
 $$
{\cal Z}_{f} = \prod_{primes}\{ 1 + p^{-\beta}\},
\eqno(18)
$$
which may be contrasted with (10). In terms of zeta functions it is:
 $$
{\cal Z}_f(\beta) = {{\zeta (\beta)}\over {\zeta (2\beta)}}.
\eqno(19)
$$
The $\beta \to 0$ limit of ${\cal Z}_f$ is evidently  quite interesting and
will be examined
elsewhere.\PP

The appearance of Heisenberg groups in the analysis of completely entangled
states is, with the
benefit of hindsight, not surprising. The essential property of such states is
to transfer
operations on one particle to its partner, and this leads to an isomorphism
between a translation
group and its dual (character group). It is this isomorphism that is the
defining property of
Heisenberg groups in general.\PP

The above discussion points to a number of lines of inquiry: On the practical
side the
connection of the physics of entangled states with prime factorization may be
of interest in
quantum computing. On the theoretical side one may try to do with entangled
states what one does
with prime numbers, i.e.\ extend the field and thereby factorize what was
previously
unfactorizable. Thus $p=2$ and primes of the form $p = 4n +1$ become non-primes
when the integers are
complexified, e.g.\ $ 5 = (2 + i)(2 - i)$. Since $p=5$ is the Hilbert space of
spin 2, it would be
 particularly interesting to discover an analagous factorization process for
$p=5$ entangled
states.
\PP \PP

\end{document}